\documentclass[prb,twocolumn,floatfix,showpacs]{revtex4}

\usepackage{graphicx}

\begin{document}

\title{Standard model for liquid water withstands x-ray probe}
\author{David Prendergast\footnote{
          Present address: University of California, Berkeley} 
        and Giulia Galli\footnote{
          Present address: University of California, Davis}
        } 
\affiliation{Lawrence Livermore National Laboratory, L-415, P.O.  Box 808, 
             Livermore, CA 94551.}

\date{\today}

% ----------------------------------------------------------------------
\begin{abstract}
% ----------------------------------------------------------------------
We present a series of ab-initio calculations of spectroscopic 
properties of liquid water at ambient conditions. Our results show that 
all available theoretical and experimental evidence is consistent with 
the standard model of the liquid as comprising molecules with 
approximately four hydrogen bonds. In particular, this model cannot be 
discounted on the basis of comparisons between measured and computed 
x-ray absorption spectra (XAS), as recently suggested. Our simulations 
of ice XAS including the lowest lying excitonic state are in excellent 
agreement with experiment and those of the TIP4P model of water are in 
reasonable agreement with recent measurements. Hence we propose that 
the standard, quasi-tetrahedral model of water, although approximate, 
represents a reasonably accurate description of the local structure of 
the liquid.
% ----------------------------------------------------------------------
\end{abstract}
% ----------------------------------------------------------------------


\maketitle

% Intro - 3 paras

Understanding the hydrogen-bonding in liquid water is fundamental to a thorough 
comprehension of the driving forces behind many physical, chemical and 
biological processes.
For the past forty years, water has been modeled as having a
local structure not dissimilar to that of the solid phase. In crystalline ice, 
each molecule is hydrogen-bonded to exactly four others, in an
approximately tetrahedral 
arrangement.~\cite{Bernal_Fowler_1933_jcp,Pauling_1935_jacs}
In the standard model of the liquid the local, molecular coordination is also 
approximately tetrahedral, with about 
3.6 hydrogen bonds/molecule.~\cite{Soper_2000_chemphys}
Classical potentials, fitted to x-ray and neutron scattering data and to 
measured thermodynamic properties, yield a quasi-tetrahedral local structure 
of the liquid when used in 
molecular dynamics (MD) simulations, and simultaneously reproduce many 
other physical properties.~\cite{Jorgensen_Chandrasekhar_1983_jcp}
More recently,
first principles, density functional 
theory~\cite{Hohenberg_Kohn_1964_pr,Kohn_Sham_1965_pr} (DFT) 
calculations have confirmed the 
quasi tetrahedral coordination of water molecules in the liquid. 
However, when neglecting proton quantum effects, 
converged DFT-MD simulations produce
a more ice-like structure for the 
liquid~\cite{Grossman_Schwegler_2004_jcp,Schwegler_Grossman_2004_jcp} 
than that inferred from experimental
estimates of the radial distribution functions.~\cite{Soper_2000_chemphys}

X-ray absorption spectroscopy techniques may help to elucidate
the local structure of liquid water.
In these approaches, high-energy, x-ray photons excite electrons
from deep atomic core levels to states above the Fermi level. 
If structural models are available, together with their electronic 
structure, it is 
possible to simulate the x-ray absorption spectra (XAS) 
of several candidate structures and establish
which one best represent the measurements.
The success of this approach hinges on the accuracy of XAS simulations
and that is the focus of this letter.

Recent XAS experiments on water have 
questioned the standard, tetrahedral model of the 
liquid.~\cite{Wernet_Nordlund_2004_science}
These experiments have compared the XAS of bulk ice, the ice surface and
water and concluded that the liquid contains significantly more broken
hydrogen-bonds than previously thought. 
The spectra are qualitatively consistent with independent 
experiments,~\cite{Bluhm_Ogletree_2002_jpcm,Smith_Cappa_2004_science} 
however, their interpretation remains controversial.
In conjunction with DFT
calculations of the XAS, the experimental results have led to the
proposal of a new structural model in which each
molecule is strongly hydrogen-bonded to only two others.
These conclusions have major implications for the physics and 
chemistry of liquid water.

In this work, we report on first principles DFT simulations of the 
XAS of ice and liquid water
at ambient pressures. Our approach can reliably predict 
the near $K$ edge stucture of oxygen in crystalline ice. 
Given the very good agreement between theory and experiment found in the 
case of ice, we used the same approach to simulate the XAS of the liquid 
as described by the TIP4P classical 
potential;~\cite{Jorgensen_Chandrasekhar_1983_jcp} 
this potential yields a local structure consistent
with the standard model. 
Our results are in reasonable agreement with experimental XAS. 
Further analysis of molecular species of the liquid with broken 
hydrogen-bonds shows that their spectral signature is qualitatively 
different from the
average XAS of the liquid; hence increasing the proportion of these 
species would yield spectra which are significantly different from the
measured ones. We also discuss previous theoretical 
approaches to simulate the XAS of 
water,~\cite{Wernet_Nordlund_2004_science,Hetenyi_DeAngelis_2004_jcp} 
and we show that our results and those present in the literature 
are all consistent with a quasi-tetrahedral model of the local structure 
of the liquid.

% Our calculations

The x-ray absorption cross section is calculated to first order using
Fermi's golden rule:
\begin{equation}
  \sigma(\omega) = 4 \pi^2 \alpha_0 \hbar \omega
                   \sum_f \left| M_{i \rightarrow f} \right|^2
                   \delta ( E_f - E_i - \hbar \omega ) \ ,
\end{equation}
where $\hbar \omega$ is the energy of the absorbed photon, which should
match the energy difference $E_f - E_i$ between initial and final electronic
states;
$\alpha_0$ is the fine structure constant;
and $M_{i \rightarrow f}$ are the matrix elements of the transition 
between initial and final states: $\left| \Psi_i \right>$ 
and $\left| \Psi_f \right>$,  evaluated 
within the electric-dipole approximation as
\begin{equation}
  M_{i \rightarrow f} = \left< \Psi_f \left| 
                        {\bf \hat{\epsilon} \cdot R}
                        \right| \Psi_i \right>
                      \approx
                        S \left< \psi_f \left|
                        {\bf \hat{\epsilon} \cdot r}
                        \right| \psi_i \right> \ ,
\end{equation}
where ${\bf \hat{\epsilon}}$ is the polarization direction of the 
electromagnetic vector potential, ${\bf R}$ and ${\bf r}$ are the 
many-electron and single-electron position operators respectively,
and $\left| \psi_{i,f} \right>$ refer to the pair of single-particle states
involved in the transition. Here the initial state is fixed
as the 1$s$ eigenstate of the oxygen atom. Single-particle
approximations of the many-electron matrix elements are accurate 
up to a factor $S$, approximately constant for all 
transitions.~\cite{Mahan_2000}

In our approach the final state of the electronic system is calculated
in the presence of the core hole which results from the x-ray excitation.
%Ideally, assuming that DFT is accurate in describing this process,
%one could
%find the self-consistent field (SCF) solution for all possible 
%x-ray excited states as a function of energy. 
%This procedure is extremely demanding, from a computational standpoint.
%Therefore we occupy only the first available empty band with the
%excited electron, and relax the electronic structure within this 
%constraint.~\cite{Mo_Ching_2000_prb}
To reduce the computational cost, we occupy only the first available 
empty band with the excited electron and relax the DFT electronic structure 
within this constraint.~\cite{Mo_Ching_2000_prb}
We then use the corresponding self-consistent potential to generate
unoccupied levels higher in energy than the first excitation.
We use the pseudopotential approximation, and 
model the x-ray excited atom with a pseudopotential derived from an oxygen atom with one electron
removed from the 1s level.
We employ norm-conserving Hamann pseudopotentials~\cite{Hamann_1989_prb}
for oxygen and a Gaussian pseudopotential for hydrogen
by Giannozzi. The
breaking of spin-degeneracy accompanying a single-electron
excitation is not included. 
The matrix elements $M_{i\rightarrow f}$ between the atomic
core level and the excited conduction band are computed using a frozen-core
approximation.~\cite{Taillefumier_Cabaret_2002_prb,
                     Hetenyi_DeAngelis_2004_jcp,
                     Bloechl_1994_prb}
%\footnote{
%  For first row elements, the dipole allowed transitions following
%  $K$ edge excitation, are to states with $p$-character. The radial
%  component of the 2$p$ atomic eigenfunction has no node and, 
%  in the core region,
%  the difference between the $p$-character of the all-electron  (AE) and 
%  pseudo wave functions is negligible. 
%  For the oxygen $K$ edge, the calculated correction to the transition matrix 
%  elements due to inclusion of AE core contributions has 
%  no visible impact on the computed XAS. 
%}
We employ DFT within the 
generalized-gradient-approximation using the Perdew-Burke-Ernzerhof
exchange-correlation functional (PBEGGA)~\cite{Perdew_Burke_1996_prl} 
and the {\tt PWSCF} code.~\cite{pwscf}. 
The single particle wavefunctions are expanded in a plane-wave basis 
with an energy cut-off of 85 Ryd.

\begin{figure}[t]
  \resizebox{\columnwidth}{!}{\includegraphics{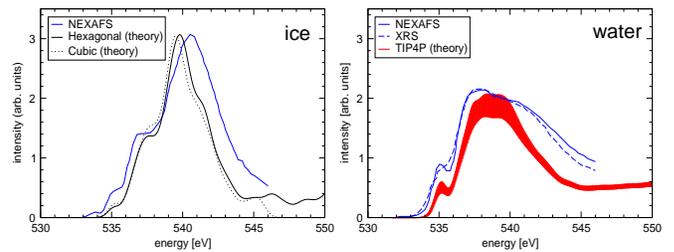}}
  \caption{{\em Left:} NEXAFS spectrum of crystalline ice I
           from Ref.~\cite{Wernet_Nordlund_2004_science}  
           (blue) and calculated XAS of heaxgonal ice I$h$
           (black,solid) and cubic ice I$c$ (black,dotted).
           {\em Right:} NEXAFS (blue,solid) 
           and XRS (blue,dashed) spectra of water from 
           Ref.~\cite{Wernet_Nordlund_2004_science},
           and calculated XAS of water (red) from the TIP4P MD
           simulation at 300~K. Vertical thickness of the calculated curve
           is the associated standard error from a sample of 32 water
           molecules.
           }
  \label{fig1}
\end{figure}

Recent NEXAFS experiments~\cite{Myneni_Luo_2002_jpcm,
Wernet_Nordlund_2004_science} reported x-ray absorption
of crystalline ice samples prepared by molecular deposition on
the Pt[111] surface.
Helium scattering indicates~\cite{Braun_Glebov_1998_prl} 
the presence of 
molecular crystals with surfaces 
consistent with either the (0001) surface of
hexagonal ice (I$h$) or the (111) surface of cubic ice (I$c$).
(Recent work~\cite{Kimmel_Petrik_2005_prl} indicates that coverage of
crystalline ice on Pt[111] may not be uniform).
In Fig~\ref{fig1} we compare the NEXAFS measurements with our calculated XAS.

We used lattice constants of ice I$h$ yielding a density of 
1.00 g cm$^{-3}$ with a $c/a$ ratio of 0.945. 
The density of ice I$c$ was fixed
at the experimental value of 0.931 g cm$^{-3}$ using $c/a$ = 1.0.
\footnote{
  The density dependence of the qualitative XAS of ice was not noticeable
  within this 7\% range. However, a full exploration of the structural phase 
  space was not carried out in this investigation.
}
The primitive cell atomic structures were relaxed, 
with these constraints on the unit cell volumes, 
until the forces computed using PBEGGA were less than 10$^{-3}$ a.u.

The primitive cell of ice I$h$ (I$c$)
contains 12 (8) water molecules and we constructed
a $2 \times 2 \times 2$ supercell of 96 (64) molecules for our calculations.
\footnote{
  Using a single $k$-point at the zone-center we found no significant 
  spectral difference in XAS computed using $2 \times 2 \times 2$
  or $3 \times 3 \times 3$ supercells for ice I$h$.
}
The excited state electronic structure, including the
excited electron in the lowest available conduction band, was
converged
using a uniform grid of 27 (64) $k$-points in the Brillouin zone
containing 10 (36) symmetry-inequivalent points. 
All absorption cross sections were averaged over the three 
Cartesian directions.

The agreement between calculated and measured XAS (see Fig. 1) is
excellent, with all qualitative features reproduced accurately.
The computed spectrum is aligned with an onset energy of 535 eV
associated with the lowest computed transition (at the zone center), 
broadened using a Gaussian lineshape with 0.4~eV 
standard deviation, and renormalized to reproduce
the peak height of experiment.
In general the width of the computed peaks is too narrow
in comparison with experiment. This is consistent with the general 
underestimation of band-width within current local approximations to DFT
and originates from an approximate description of exchange interactions. 
Improvements are
possible using self-interaction corrected DFT 
calculations~\cite{Perdew_Zunger_1981_PhysRevB} or
by using $GW$ quasi-particle corrections.~\cite{Hybertsen_Louie_1986_prb}
Note that ice I$h$ and I$c$ differ structurally only beyond
first nearest neighbor molecules.
Consequently, their spectra are very similar, given the short spatial range of
the oxygen $K$ edge transition probability, 
which is proportional to the overlap of the $p$-character of
conduction bands with the oxygen 1$s$-orbital, 
possessing a 0.2~\AA~effective radius.

\begin{figure}[t]
  \resizebox{\columnwidth}{!}{\includegraphics{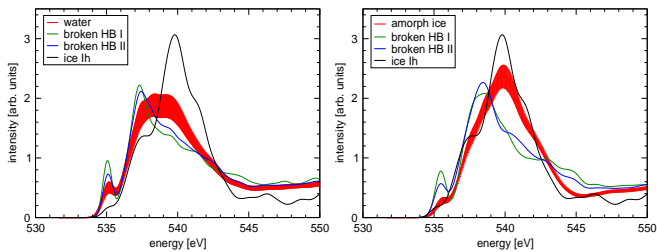}}
  \caption{{\em Left :} Comparison of the calculated XAS of water (red) and 
           crystalline ice I$h$ (black).
           {\em Right:} A similar comparison between ice I$h$ and
           a model of amorphous ice (red).
           Also indicated in both panels are the characteristic spectra from
           water molecules with at least one broken donor hydrogen bond 
           (see text).
           Vertical thickness indicates standard error as in Fig.~\ref{fig1}. 
           We assume that the experimental 
           data for solid and liquid~\cite{Wernet_Nordlund_2004_science}
           are quantitatively comparable 
           and use the same normalization factor and
           onset energy for both theoretical curves.
           }
  \label{fig2}
\end{figure}

% Our results for water
We model liquid water using the classical, TIP4P point charge 
model,~\cite{Jorgensen_Chandrasekhar_1983_jcp} 
which qualitatively reproduces the phase
diagram of water,~\cite{Sanz_Vega_2004_prl} and 
which we have previously used to analyze the electronic
structure of the liquid.~\cite{Prendergast_Grossman_2005_jcp} 
In this model water molecules are treated as rigid.
We took 10 snapshots of 32 water molecules
from a 200~ps MD simulation~\cite{gromacs1,gromacs2} spaced 20~ps apart.
To approximate the XAS experiment,
we average computed spectra from each molecule in a given snapshot and 
further average over all snapshots. 
In total we combined the results of 320 spectral calculations to 
approximate the XAS of the liquid.
We found very little variation between the XAS of individual, uncorrelated 
snapshots computed using only the zone-center $k$-point.
Therefore, we picked one representative snapshot of 32 molecules, 
and carried out calculations using 27 $k$-points in the Brillouin Zone, 
for each excited molecule.
Each spectrum of discrete transitions is broadened with a 
Gaussian lineshape with a standard deviation of 0.3~eV.
The averaged XAS is compared in Fig.~\ref{fig1} 
with the experimental NEXAFS and XRS spectra of water at 
300~K.~\cite{Wernet_Nordlund_2004_science}
The vertical width of the theoretical curve indicates an estimate of the
standard error at a given energy.
Our calculation is in good qualitative agreement with experiment, 
indicating that the tetrahedral local model of the liquid accounts 
reasonably well for measured XAS.
%\footnote{ 
%  The 32 molecule supercells of our calculations
%  may introduce finite size errors with respect to
%  containment of the first excited electron. 
%  However, the XAS of ice I$h$ is converged using a 96 molecule supercell.
%  Assuming that
%  the dispersion of a band is proportional to $V^{-2/3}$ ( $V$ is
%  the supercell volume), the dispersion of the occupied
%  excited electron band in a
%  32 molecule supercell of ice is estimated to be 0.40~eV; this coincides with 
%  the calculated, average dispersion of (0.41$\pm$0.10)~eV 
%  in the excited electron band of the liquid. Therefore, we might
%  not expect large differences for the XAS of the 32 molecule liquid supercell
%  with respect to a fully converged calculation
%}

The contribution to the XAS of the liquid from molecules with broken or
distorted hydrogen-bonds (HB) is compared to that of the entire liquid in
Fig.~\ref{fig2}. Our model contains approximately 20\% of molecules 
with broken HB,
which are defined according to two definitions: 
(I) exceeding a maximum separation of 
oxygen atoms of 3.5~\AA~ and a maximum angle of the donated hydrogen 
from the line joining both oxygens of 40$^\circ$; 
(II) exceeding
a maximum oxygen-oxygen separation which depends on the angle
as outlined in Ref.~\cite{Wernet_Nordlund_2004_science}.

There is a clear,qualitative difference between the spectra of
species with broken HB and the averaged spectrum. 
Such differences were predicted in 
Ref.~\cite{Wernet_Nordlund_2004_science}, 
where it was concluded that the liquid should contain about 80\% 
of broken HB species in order to produce the measured XAS.
We do not find this to be necessary when using our approach for 
the calculation of XAS, which differs from that used in 
Ref.~\cite{Wernet_Nordlund_2004_science} . 

%Ice and amorphous.
In Fig.~\ref{fig2}  we also compare the XAS of 
ice I$h$ and of a rather primitive model of amorphous or disordered ice,
generated using the TIP4P potential
by quenching a 300~K liquid sample of 32 molecules down to
100~K. One snapshot at this temperature was used to generate
the XAS in Fig.~\ref{fig2}. The spectra of the broken HB 
species are qualitatively similar in the disordered sample and in 
the liquid.
However, only one (two) of the
molecules in this snapshot had less than two donor HB according to
definition I (II), and yet the 
XAS is qualitatively different from that of the 
crystalline sample.
Clearly disorder plays a role in reducing the peak height at 539--540~eV.
We also observe an increase in intensity around 538~eV and at much
higher energies above 545~eV. 
Both of these trends are seen in the liquid as well,
and can be associated with disorder in the oxygen sublattice.
On the other hand, the significant increases in intensity at the onset (535~eV) 
and the main-edge (537-538~eV) found in the liquid can be associated with 
broken HB,
of which there are approximately 4--6 times as many in the liquid sample. 
There are therefore two important effects accounting for the difference 
between XAS of ice and XAS of the liquid: broken HB and disorder of the 
oxygen sublattice.

% Previous approximations

\begin{figure}[tb]
  \resizebox{\columnwidth}{!}{\includegraphics{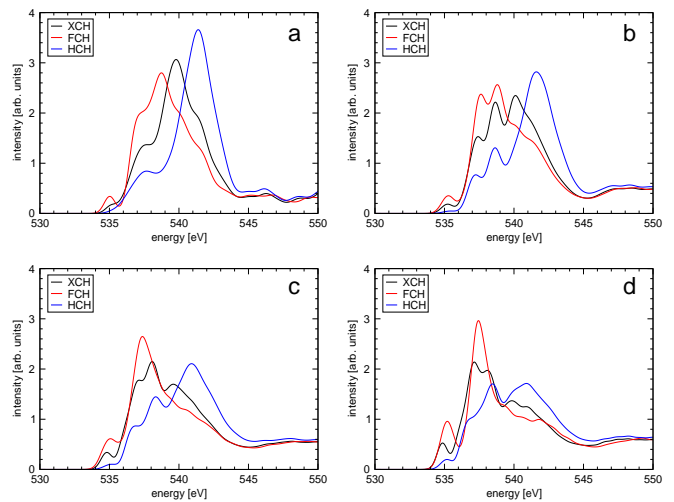}}
  \caption{Comparison of DFT XAS within various
           approximations for the x-ray absorption process, {\it viz.}
           excited state core hole (XCH), full core hole (FCH) and
           half core hole (HCH). (a) hexagonal ice; (b) amorphous ice;
           (c) liquid water; (d) liquid water sampling only
           broken-hydrogen-bonded species.
           Note that the calculated 
           HCH spectrum of a standard, quasi-tetrahedral liquid water model
           appears ice-like in comparison with experiment,
           however it is still clearly distinguishable from ice XAS.
          }
  \label{fig3}
\end{figure}

In the recent literature, two approaches have been applied
to the simulation of the XAS of water in various phases: 
the full core hole (FCH) and half-core hole (HCH) approaches. 
The FCH technique ~\cite{Hetenyi_DeAngelis_2004_jcp} 
is analogous to the one adopted
here, except that there is no self-consistent inclusion of the excited
electron. Such an
approach should accurately reproduce the higher energy excitations to
progressively more delocalized states, and at high energy yield the same 
results as our approach, which we call excited-state-core-hole (XCH). However, accounting for 
the impact of a localized excited electron at low energies leads to
differences between FCH and XCH results; these are shown in 
Fig.~\ref{fig3}. FCH tends to overestimate the absorption
cross section at and near the onset, consequently underestimating the
main peak height, by virtue of the oscillator strength sum-rule.

In the HCH approach only half of an electron is removed from the core-state,
thereby simulating a transition state in the x-ray excitation process.
We have found that in the presence of the HCH pseudopotential,
inclusion of an excited electron at the conduction band minimum has almost
no impact on the calculated XAS. The HCH spectra tend to 
overemphasize the main peak and underestimate the excitonic
near-edge intensity. This is consistent with the reduced binding energy
of the half-core-hole in the excited oxygen atom.

We note that the same spectral trends exist in all approaches (FCH, XCH and HCH) with respect to 
increasing disorder and number of broken HB: a decrease in
the main peak height and an increase in intensity at the near-edge are observed.
However, in the limit of converged $k$-point sampling and when employing
homogeneous numerical broadenings, we find that only the XCH approach
provides a consistent agreement with experiment.

% ------------------------------------------------------------------------
% Conclusions
% ------------------------------------------------------------------------

In summary, using DFT electronic structure calculations, 
we find
excellent agreement between the calculated and experimentally measured
XAS of ice.
Using the TIP4P classical potential to simulate a standard, quasi-tetrahedral
model of liquid water we find reasonable
agreement between our DFT calculations and experiment.
Furthermore, restriction of our analysis to those molecular species with broken 
HB leads to simulated XAS which are qualitatively different from
experiment. Therefore we conclude that the percentage of broken HB
in the standard model is consistent with experimental observations.
Finally, despite the different spectra produced by various
approximations used to simulate the x-ray absorption process, 
we find no conclusive evidence within each approach
to discount the standard model of the liquid. 

% ------------------------------------------------------------------------
\begin{acknowledgments}
% ------------------------------------------------------------------------

We wish to acknowledge E. Schwegler, T. Ogitsu, G. Cicero, F. Gygi, and 
A. Correa for useful discussions. 
This work was performed under the auspices of the U.S. Department of Energy
at the University of California/Lawrence Livermore National Laboratory under
Contract No. W-7405-Eng-48.

\end{acknowledgments}

%\bibliography{strings,ice,water,dft,books,xas_theory,methods,codes}

\end{document}